\documentclass[aps,prb,twocolumn,reprint,superscriptaddress,showkeys,floatfix,amsmath,amssymb]{revtex4-2}
\usepackage{epsfig}
\usepackage[latin9]{inputenc}
\usepackage{color}
\usepackage{booktabs}
\usepackage[pdftex,colorlinks=true,linkcolor=red,citecolor=blue]{hyperref}
\usepackage{graphicx}
\usepackage{hypernat}
\usepackage{csquotes}
\usepackage{xcolor, soul}
\usepackage{dcolumn}
\usepackage{bm}
\newcommand{\rtcom}[1]{\hl{}}{}

\begin{document}

\preprint{APS/123-QED}

\title{Modulation of electronic and piezoelectric properties  of lead-free halide perovskites LiSnX$_3$ (X = Cl, Br, and I) under applied pressure}

\author{Celestine Lalengmawia}
\affiliation{Advanced Computation of Functional Materials Research Lab (ACFMRL) Department of Physics, Mizoram University, Aizawl-796004, India}
\affiliation{Physical Sciences Research Center (PSRC), Department of Physics, Pachhunga University College,  Aizawl-796001, India}
\author{R. Zosiamliana}
\affiliation{Advanced Computation of Functional Materials Research Lab (ACFMRL) Department of Physics, Mizoram University, Aizawl-796004, India}
\affiliation{Physical Sciences Research Center (PSRC), Department of Physics, Pachhunga University College,  Aizawl-796001, India}
\author{Bernard Lalroliana}
\affiliation{Advanced Computation of Functional Materials Research Lab (ACFMRL) Department of Physics, Mizoram University, Aizawl-796004, India}
\affiliation{Physical Sciences Research Center (PSRC), Department of Physics, Pachhunga University College,  Aizawl-796001, India}
\author{Lalhum Hima}
\affiliation{Advanced Computation of Functional Materials Research Lab (ACFMRL) Department of Physics, Mizoram University, Aizawl-796004, India}
\affiliation{Physical Sciences Research Center (PSRC), Department of Physics, Pachhunga University College,  Aizawl-796001, India}
\author{Shivraj Gurung}
\affiliation{Physical Sciences Research Center (PSRC), Department of Physics, Pachhunga University College,  Aizawl-796001, India}
\author{Lalhriat Zuala}
\affiliation{Physical Sciences Research Center (PSRC), Department of Physics, Pachhunga University College,  Aizawl-796001, India}
\author{Lalmuanpuia Vanchhawng}
\affiliation{Physical Sciences Research Center (PSRC), Department of Physics, Pachhunga University College,  Aizawl-796001, India}
\author{Amel Laref}
\affiliation{Department of Physics and Astronomy, College of Science, King Saud University, Riyadh, 11451, Saudi Arabia}
\author{A. Yvaz}
\affiliation{World-class research center "Advanced Digital Technologies", State Marine Technical University, Saint Petersburg, 190121 Russia}
\author{D. P. Rai}
\email[D. P. Rai:]{dibyaprakashrai@gmail.com}
\affiliation{Advanced Computation of Functional Materials Research Lab (ACFMRL) Department of Physics, Mizoram University, Aizawl-796004, India}

	\date{\today}

\begin{abstract}
Pb-based perovskites are considered to be the most efficient materials for energy harvest. However, real-time application is limited because of their toxicity. As a result, lead-free perovskites that offer similar advantages are potential alternatives. Here, we have chosen LiSnX$_3$ (X = Cl, Br, and I) for further calculation and explore its possibilities for harvesting clean and green energy. Our objective is to examine strategies for optimizing the parameters that control the energy-harvesting capabilities, particularly the interplay between structural variations and electrical properties. The density functional theory (DFT) has been employed for the theoretical simulation. Within the DFT framework, we have studied the effect of applied pressure (0 to 20 GPa) and elemental substitution on their physical properties. We hereby report the variation of lattice parameters, elastic constants, band gaps, and piezoelectric constants. MD simulation with time steps of up to 5 ps was performed to verify structural stability at room temperature. We report the semi-conducting characteristic of LiSnX$_3$ and the high piezoelectric response up to 20.7 Cm$^{-2}$. The presence of high piezoelectric coefficients suggests that manipulation of the structure of LiSnX$_3$ may provide an alternative way to harvest energy through electromechanical processes.

\end{abstract}

\maketitle


\section{Introduction}
The global increase in energy demand due to overconsumption is directly related to overpopulation and various ambitious development projects. Hence, it is alarming for the sustainability of life in the near future and urges us to look for alternative ways to generate and store energy without disrupting the ecosystem\cite{Ang2022} Scientists and engineers have opted for various strategies to gather and produce energy\cite{Qazi2019}. Manipulation of the energy transfer mechanism by absorbing electromagnetic solar radiation through solar cells is the most common, clean, and green way to harvest energy.\cite{Celestine2024d} Even though solar panels are the best means to generate energy, they have significant drawbacks. The low efficiency of solar panels for mass use, their large size, inability to store energy, and the tainted electrodes due to exposure to the environment are just a few of the problems. Another major concern is the disposal of solar panels. \cite{Rajagopal2017a, Rahaman2018d, Schwarz1996d}. The other factor being the simultaneous production and storing of energy is a big issue. There are several well-developed alternatives that could produce energy, such as hydro \cite{Azad2020a}, thermoelectricity\cite{Jaziri2020}, piezoelectricity\cite{Mahapatra2021}, wind \cite{Ibrahim2011a}, etc., which depend on the influence of external fields such as pressure, temperature, strain, etc. Another major source of energy is nuclear energy, which requires huge amounts of nuclear fuels, produces a large amount of nuclear by-products that are hazardous to the environment and accidental failure could lead to disaster.\cite{Rajan2018} On the other hand, thermal energy \cite{Luo2015}, requires the burning of fossil fuels to generate energy and contributes to environmental pollution by emission of greenhouse gases. \cite{Filonchyk2024} Among these energy harvesting techniques, piezoelectricity is one such method in which materials convert mechanical stress into electrical energy. This technique looks promising owing to its durability, adaptability and novelty in the creation of green renewable energy. The conversion of energy by piezoelectricity requires little effort, as mechanical sources such as hydraulic, fluid, vibrational, and acoustic waves are readily available and have a negligible impact on the local environment. \cite{Jeong2017b} Though the mechanical sources are easily available, finding an efficient material that converts them into energy is a very challenging task. From our rigorous literature survey, we have come across several materials, but the materials with noncentrosymmetric symmetry having large band gaps and high dielectric polarization are crucial for this kind of research. Piezoelectric materials promise a variety of applications, such as nano-sensors\cite{Lim2006}, nano-robots\cite{Chen2023}, Microelectronic mechanical systems (MEMS)\cite{Liu2011}, piezoelectric force transducers\cite{Mihai2011}, Piezoelectric actuators\cite{Aabid2023}, Piezoelectric buzzers\cite{Mishra2016},  Piezoelectric humidifiers\cite{Putra2020}, smart bulbs\cite{Niu2020}, etc.\cite{Xue2021c}. 

\par The underlying physics of piezoelectricity is that when an external force is applied to a material, an electric dipole moment polarizes and generates electrical energy\cite{Elahi2018}. This phenomenon was first discovered by the Curie brothers in 1880\cite{Martin1972}. Several substances have since been confirmed to be piezoelectric, including crystals, polymers, biomolecules, and two-dimensional layered materials. Piezoelectric effects are most prominent in noncentrosymmetric structures\cite{Celestine2024e}. As mentioned earlier, research has been conducted on multiple compounds, and it has been found that some materials give a high or decent piezoelectric response. Until now, a wide range of piezoelectric materials have been synthesized and investigated, including glass-like substance (Na$_2$SiO$_3$) with 0.10 Cm$^{-2}$, zinc sulfide (ZnS) produced around 0.16 Cm$^{-2}$, and lead zirconate titanate (PZT) around 0.31 Cm$^{-2}$  \cite{Ferahtia2014a,Chaplya2001,Zosiamliana2022m}.

\par Moreover, advances in technologies enable scientists to create piezoelectric materials with halide perovskites, which have a perfect combination of chemical and physical properties. Liu \textit{et al.}\cite{Liu2016} employing the first principles method, investigated the response of piezoelectric on methylammonium lead iodide perovskite (MAPbI$_3$) using the molecular orientation technique by measuring the interactions between polarization and strain in two model objects that have polarity and non-polarity and produced a piezo-response of 0.83 Cm$^{-2}$. Here, we also note one oxide perovskite, PbTiO$_3$, with which Li \textit{ et al.}\cite{Li1993} experimented using the flux solution method and produced a high piezoelectric coefficient of 3.35 Cm$^{-2}$. Despite having a high piezo-response, these compounds contain the toxic element (Pb) which is hazardous to humans and the environment. An alternative way to conduct an environmentally friendly process has been conducted for years, which includes the replacement of Pb with a less toxic or environmentally less destructive element. In this work, we selected compounds with free Pb-containing elements for a green and clean energy harvest approach.

\par Another fascinating study that has been used in recent years to improve the properties of a material is by inducing pressure\cite{Seo1998a}. This approach can alter the chemical and physical attributes of a material, and thus sometimes give a higher response in various properties than that of the pristine state of a compound. Therefore, a fundamental understanding of modifications in the properties of materials can be obtained by a pressure-induced study\cite{Liu2019b}. To the best of our knowledge, no research has been done on the (piezo)electromechanical characteristics of these compounds based on DFT under applied pressures.

\par For a single perovskite, a chemical formula of ABX$_3$ has been used where the cation at the A site is allotted with lithium (Li$^+$), tin (Sn$^{+2}$) as the cation at the B site, and the X site of the anion with halogens (Cl, Br, and, I). In this paper, we have calculated and analyzed various properties of the selected eco-friendly halide perovskites. Moreover, we examined the influence of pressures on these materials and how they affect the responses of properties, especially piezoelectric properties.

\section{Computational Details}
\par The implementation of density functional theory (DFT)\cite{Nityananda2017a} in QuantumATK Q-2019.12 was used for all calculations in the present study\cite{Smidstrup2019b}. This computational technique relies on the linear combination of atomic orbitals (LCAO) approach\cite{Schlipf2015b}. The generalized gradient approximation (GGA) has been utilized for electron-ion interactions within the Perdew-Burke-Ernzerhof (PBE) functional\cite{Perdew1996d}. The optimization algorithm known as Limited Memory Broyden-Flecther-Goldfarb-Shanno (LBFGS), taken from Newton-quasi methods\cite{Zhao2021a}, has been adopted to perform complete cell optimization. Consequently, throughout geometry optimization, we have relaxed all the parameters: atomic positions, cell volumes, and space groups. The criteria for Hellmann-Feynmann force and stress tolerance were established at 0.01 eV \AA$^{-1}$ and 0.00001 eV \AA$^{-3}$, respectively, with a maximum step size of 0.2 \AA and a maximum number of 200 convergence steps to determine structural and cell energy convergence. A medium base set of PseudoDojo potential\cite{VanSetten2018b}, equivalent to double zeta polarized (DZP) for Li, Sn, Cl, Br, and I, has been adopted.
\par For each compound, X = Cl, Br and I, an isotropic pressure of 0 GPa to 20 GPa, with a difference of 5 GPa, was applied to see the appearance of changes in the compounds. From pristine to pressurized states, we have sampled a Monkhorst-Pack scheme\cite{Monkhorst1976c} 8$\times$8$\times$8 k-mesh points for geometry optimization, while a denser k-point of 12$\times$12$\times$12 with the same convergence criteria has been adopted for property calculations. To verify the internal energy stability, we employed the formation energy calculations using the following equation,
\begin{equation}
E^f(LiSnX_3) = \dfrac{[E_{tot} - E_{Li}-E_{Sn}-3E_{X}]}{N}
\end{equation}
\par where E$_{tot}$ is the total energy, E$_{Li}$ represents the total energy of Lithium atom, E$_{Sn}$ is the total energy of Tin atom and E$_{X}$ are the energies of halogen atoms. Negative values of E$^f$ indicate the stable nature of the compounds. Here, tabulated in Table \ref{Table 1}.
\section{Results and Discussion}
\subsection{Structural Properties}

\par  In this work, we examined the optimal structures of the halide perovskites LiSnX$_3$ (where X = Cl, Br and I) that adopt hexagonal symmetry in their pristine state. All structural information is obtained from the online open access materials database "Materials Project"\cite{materials}. The optimized lattice parameters under the influence of isotropic pressures are tabulated in Table \ref{Table 2}. The role played by the induced pressures highlighted changes in their structural part, which could potentially impact the properties calculation results. 

\begin{figure}[hbt!]
	\centering
	\includegraphics[height=4cm]{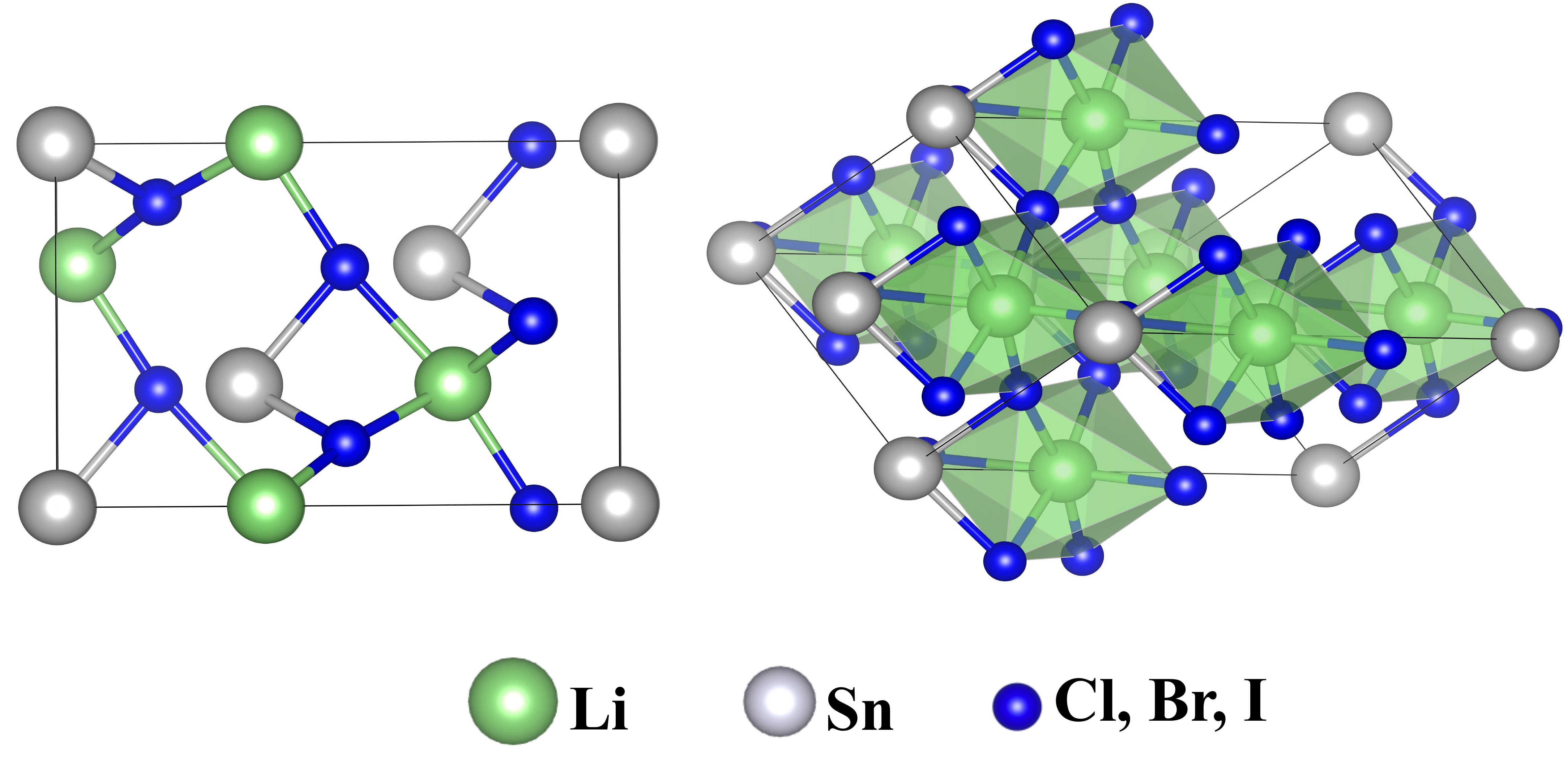}
	\caption{Schematic 2D and 3D diagrams of LiSnX$_3$ using VESTA program\cite{Momma2008b}}
	\label{Fig.1 Structures}
\end{figure}

\begin{figure}[hbt!]
	\centering
	\includegraphics[height=5.5cm]{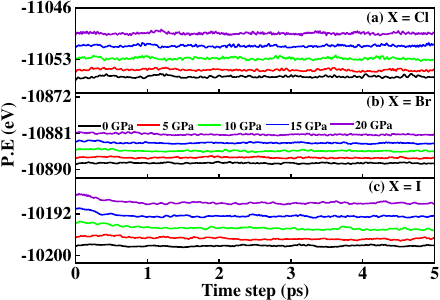}
	\caption{Calculated potential energies from MD-simulation for LiSnX$_3$ (a)X = Cl, (b) X = Br, and (c) X = I}
	\label{Fig.5 MD}
\end{figure}

\par It is a typical procedure to utilize Goldschmidt's tolerance factor "t" to assess the stability and crystal formation of 3D halide perovskites\cite{Bartel2019}.

\begin{equation}
t=\frac{r_A+r_X}{\sqrt{2}(r_B+r_X)}
\label{Eq 2}
\end{equation}

\par where r$_A$, r$_B$, and r$_X$ represent the ionic radii\cite{Verma2008,Shannon1976} of atoms A, B, and X, respectively. The tolerance factor values depend on the formation of the crystalline structure by each of the contributing atoms. Using Eq.\ref{Eq 2}, the tolerance factors of LiSnX$_3$ for their respective pristine states are calculated which are given in Table \ref{Table 1}.

\begin{table}[hbt!]
	\small
	\caption{\ Calculated ionic radii (\textit{r}) in \AA and tolerance factor (\textit{t}) of pristine LiSnX$_3$ (X = Cl, Br, and I)} 
	\label{Table 1}
	\begin{tabular*}{0.48\textwidth}{@{\extracolsep{\fill}}l|l|l|l|l}
		\hline
		Atom(s) & \textit{r} & Compound(s) & \textit{t} & Other's result\\
		\hline
		Li$_{(A)}$     & 1.61$^a$  & LiSnCl$_3$ & 0.81 & 0.91 \cite{RatulHasan2024}\\
		Sn$_{(B)}$     & 0.69$^a$   & & & \\
		Cl$_{(X_{1})}$  & 1.67$^b$  & LiSnBr$_3$ & 0.81 & 0.92 \cite{RatulHasan2024}\\
		Br$_{(X_{2})}$ & 1.82$^b$  & & & \\
		I$_{(X_{3})}$  & 2.06$^b$  & LiSnI$_3$  & 0.80 & 0.93 \cite{RatulHasan2024}\\
		\hline
	\end{tabular*}
$^a$ Ref\cite{Verma2008}, $^b$ Ref\cite{Shannon1976}
\end{table} 

\begin{figure*}[hbt!]
	\centering
	\includegraphics[height=14cm]{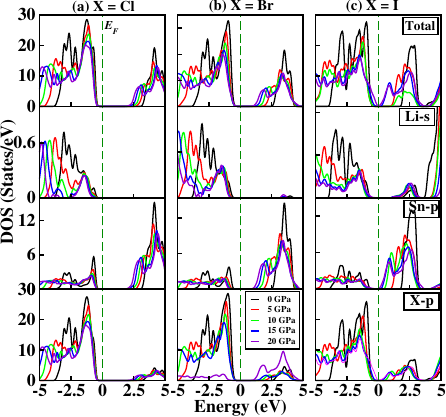}
	\caption{Calculated LiSnX$_3$(X=Cl, Br and I) Density of States. Fig(a) for X =Cl, Fig(b) for X = Br and Fig(c) for X = I}
	\label{Fig.3 DOS}
\end{figure*}

\begin{table*}[hbt!]
	\small
	\caption{\ Calculated lattice parameters, electronic band gaps (in eV), volumes, and formation energies of LiSnX$_3$ (X = Cl, Br, and I) under different pressures.}
	\label{Table 2}\renewcommand{\arraystretch}{1.25}
	\begin{tabular*}{\textwidth}{@{\extracolsep{\fill}}l|llll}
		\hline
		&(a,c)&E$_g$&V$^o$&E$^f$      \\
		\hline
		\ \ Pressure&   & X=Cl&&\\
		\hline
		\ \ \ \ \ \ \ 0 &7.02, 9.18&3.81& 409.5 &-1.85  \\
		\ \ \ \ \ \ \ 5 &6.63, 8.44&3.30& 321.8&-1.83   \\
		 \ \ \ \ \ \ 10  &6.41, 8.05&2.97&286.1 &-1.82\\
		 \ \ \ \ \ \ 15 &6.25, 7.80&2.73& 263.8&-1.88\\
		 \ \ \ \ \ \ 20 &6.14, 7.60&2.51&248.6 &-1.80\\
		Other's work & (cubic)5.60\cite{Pakravesh2024}&1.35\cite{Ghani2024a},3.38\cite{Park2019a}& &\\
		\hline
		\ \ Pressure &  &X=Br& &\\
		\hline
		\ \ \ \ \ \ \ 0 &7.58, 10.12&3.19 & 504.9&-1.42  \\
		\ \ \ \ \ \ \ 5 & 7.01, 8.97&2.41&382.8 &-1.37  \\
		 \ \ \ \ \ \ 10  &6.76, 8.51&1.97& 337.7&-1.32\\
		 \ \ \ \ \ \ 15 &6.59, 8.23&1.65& 310.5&-1.28\\
		 \ \ \ \ \ \ 20 &6.47, 8.02 &1.42& 291.3&-1.25\\
		Other's work &(cubic) 5.72\cite{Pakravesh2024}&2.7\cite{Park2019a}& &\\
		\hline
		\ \ Pressure&   &X=I& &\\
		\hline
		\ \ \ \ \ \ \ 0&8.11, 10.69&2.42& 610.1&-0.99\\
		\ \ \ \ \ \ \ 5 &7.54, 9.55&1.56& 470.4&-0.93\\
		 \ \ \ \ \ \ 10  &7.23, 9.03&0.99& 409.3&-0.84\\
		 \ \ \ \ \ \ 15 &7.03, 8.73&0.57&374.3 &-0.77\\
		 \ \ \ \ \ \ 20 &6.89, 8.51&0.11&350.8 &-0.83\\
		Other's work &(cubic)5.80\cite{Pakravesh2024}&2.27\cite{Park2019a}& &\\
		\hline
	\end{tabular*}
\end{table*}

\par The cells' geometry optimizations have been achieved from pristine to pressurized states at their respective minimal energies. The lattice parameters and volumes for these compounds under different pressures are tabulated in Table \ref{Table 2}. We further report that in Table \ref{Table 2}, substituting halogen atoms from Cl$\rightarrow$Br$\rightarrow$I increases the lattice constant \cite{Wang2016a} while inducing pressures decreases it\cite{Khan2023a}. As a result, decreasing lattice constants under hydrostatic pressure imply that the interatomic distance (bond length), which indicates a strong interaction between the atoms, may gradually decrease.

To validate and analyze atoms and molecular mobility, we performed Molecular Dynamics (MD) simulations\cite{Brooks1989a} within a canonical ensemble incorporating QuantumATK's Nose Hoover thermostat\cite{Martyna1992c}. Since the 1960s, its application has grown in popularity and has been used with both ordered and disorganized solids, liquids, and gases. When applied to solids, it has the unique benefit of organically including structural relaxations and anharmonic forces\cite{Goncalves1992}. This technique can be used to determine the physical motions and trajectories of the system's atoms, molecules, and nanoparticles. By presuming that particles have an interaction potential, it can forecast the physical characteristics of materials. Various potentials and force fields describe how atoms interact with each other. Because MD does not make any additional fundamental assumptions, it can be a reliable instrument for examining the outcomes of other classical models with a broad range of applications in the domains of nanotechnology, biochemistry, and biophysics.

\par The evolution of potential energies (PE) in 5-ps time steps for the investigated eco-friendly perovskites is depicted in Fig.\ref{Fig.5 MD}. Herein, we employed the nVT-based canonical ensemble. During these simulations performed at 300K, the number of particles, volumes, and temperature was made constant to acquire comprehensible results in the evolution of the energies. The PE's in Fig.\ref{Fig.5 MD} represent the overall energy generated by the interplay between the systems' bonding and non-bonding \cite{Yu2020a}. Since the PE's simulated profiles show nearly linear fluctuations, the systems are assumed to be thermally stable.

\par Apart from these generated potential energies, we have also obtained the vibrational density of states (VDOS) and specific heat capacities (\textit{C$_v$}) of the mentioned halide perovskites, which are depicted and discussed in S1 \& S9. 

\subsection{Electronic Properties}
\par When researching atomic-level interactions in any material, its electrical properties remain one of the most important things to consider\cite{Renthlei2023f}. By replacing the halogen atoms down the group, from Cl$\rightarrow$Br$\rightarrow$I, the compounds' bandgaps reduce in their pristine states because the increase of the atomic orbitals weakens the interaction between the nucleus of an atom and its outermost electron. Since the chlorine atom (Cl) is present, LiSnCl$_3$ has the highest electronegativity of the three compounds and the largest band gap. In contrast, LiSnI$_3$ has the smallest bandgap among the three because of the low electronegativity of the I atom\cite{Celestine2024e}. As pressure increases from 0 GPa to 20 GPa, each compound experiences a decrease in the band gaps. (See Table \ref{Table 2}). The band gaps obtained from the pristine compounds were then compared with the existing theoretical data.

\par Fig.\ref{Fig.3 DOS} shows the density of states (DOS) of the compounds studied. This study illustrates the level of contribution for each atomic orbital in the exchange of photons from the valence to the conduction bands. Here, for all DOS, from pristine to pressurized states, we have taken an energy range from -5 eV to +5 eV. The valence orbital of each atom has been taken into consideration, i.e. the s-orbital of Lithium atom, the p-orbitals of Tin and the p-orbitals of Chlorine, Bromine, and Iodine. In the figures, we can see that the contributions of the halogen atoms are higher in the valence bands, while the contribution of the tin (Sn) atom is higher in the conduction bands. Further observation reveals that, when pressures were exerted from 0 to 20 GPa, the band gaps decrease with the amplitude of the orbital contributions. Hence, these pressure-induced studies show that the compounds under investigation are potential and promising candidates for various applications including power capacitors for wider band gaps, optoelectronics, and piezoelectric applications for decent and narrow bandgap materials.

\subsection{Electron Density Difference}

\par
\begin{figure*}[hbt!]
	\centering
	\includegraphics[height=14cm]{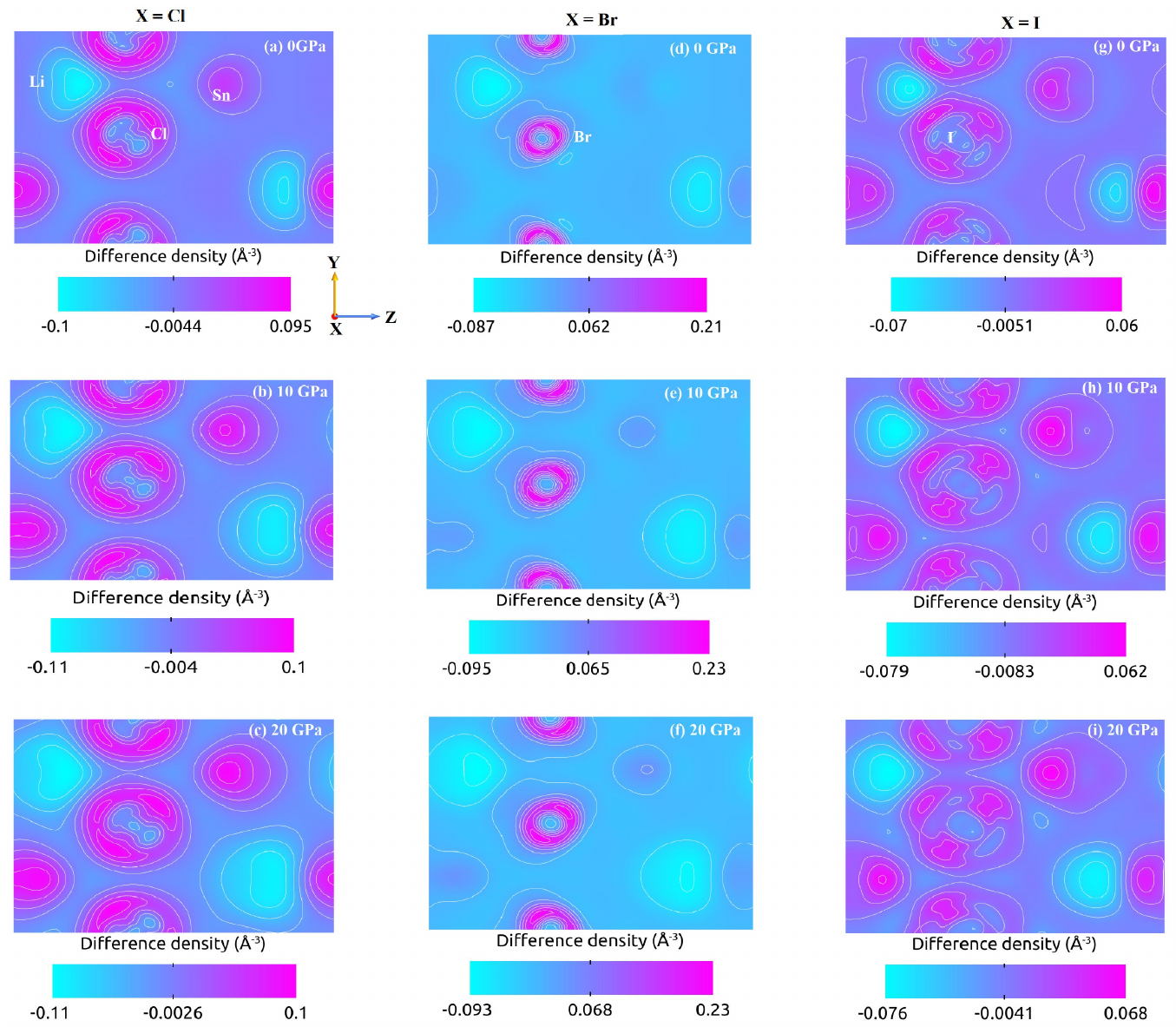}
	\caption{Calculated Electron Density Difference of LiSnX$_3$(X=Cl, Br and I) showing the differences of 0 GPa, 10 GPa and 20 GPa. Fig(a-c) for X =Cl, Fig(d-f) for X = Br and Fig(g-i) for X = I}
	\label{Fig.4 EDD}
\end{figure*}

 Comparing the electron densities of atoms in materials is another important function of the electron difference density (EDD), which is essential to chemical bonding\cite{Harrison2003}. EDD offers important details regarding the electronic structures and charge distributions among the atoms Li, Sn, Cl, Br, and I. Atomic recombination inside the system will result in charge transfer, which yields intriguing results. The 2D plots shown in Fig.\ref{Fig.4 EDD} show the differences in electron densities for the investigated lead-free halide perovskites from 0 GPa to 20 GPa. Here, we have shown for 0 GPa, 10 GPa, and 20 GPa for X = Cl, Br, and I, while the remaining are kept in S13. Our output figures show that the magenta-colored regions show higher electron densities, while the light blue-colored portions show lower electron densities. This indicates that higher electron-accumulated regions tend to have stronger chemical bonds, while lower-density regions have weaker bonds. Further observation reveals that increasing pressure alters the atom's electron clouds or the contour lines, which results in the distribution of charge within the material. Another finding obtained is that during the charge-transferring process, depletion of charges are experienced by the lithium (Li) and tin (Sn) atoms while the halogen (X = Cl, Br, and I) atoms accumulate the depleted electrons by their neighboring atoms. This transfer of charges influences the material's properties, including the electronegativity, electrical conductivity, and piezoelectric properties of the compound, where the latter will be discussed in one of the following sections.

\subsection{Elastic constants and Mechanical Properties}

\par Mechanical stability is a prerequisite for comprehending the true properties of a material, and we use finite-strain theory\cite{Steddon1954} to compute elastic constants (C$_{iv}$) to ensure the stability of the compound.  These elastic constants provide information about the strength of a material and reaction to an applied external force. The systems under investigation crystallize in hexagonal symmetry\cite{Yu2018c}, according to structural analysis. For the system to be mechanically stable, it should meet the following criteria:
\begin{equation}
C_{44} > 0, C{_{11}^{2}} > C{_{12}^{2}}, (C_{11} + 2C_{12})C_{33} > 2C{_{12}^{2}}
\label{Eq 8}
\end{equation}

\par From table \ref{Table 3}, it is certain that the elastic constants (C$_{iv}$) of these halide perovskites experienced an immense change when an external pressure was exerted on them. Meanwhile, the compounds in the study fulfill the Born stability criteria aforementioned \cite{Born1940c} which means that the materials, from pristine to pressurized, are mechanically stable. The values obtained for C$_{iv}$ are listed in Table \ref{Table 3} and in Tables S2, S3, and S4. We can determine a number of important mechanical properties by using elastic constants. In this context, we compute the elastic moduli, which are listed in Table \ref{Table 3} as the shear modulus (\textit{G}), bulk modulus (\textit{B}), Young's modulus (\textit{E}), and Poisson's ratio (\textit{v}). Consequently, we also observed that all bulk moduli (\textit{B}) are higher than shear moduli (\textit{G}), indicating that the compounds are more resistant to axial compression than to shear deformation. This further demonstrates that the halide perovskites under investigation are highly anisotropic. As pressure increases from 0 GPa to 20 GPa, the lattice parameters decrease for the studied perovskites, which ultimately reveals the effect of pressure.\cite{Lu2017a}

\par Making use of the elastic moduli of the compounds in Table \ref{Table 3}, we can obtain the Pugh's ratio (\textit{k}) \cite{Pugh1954} or \textit{B}/\textit{G}, which determines the failure mode of the material. A material can be considered ductile (brittle) when the value of \textit{k} is above (below) the critical value of 1.75. Finding a material's failure mode can also be done by computing Poisson's ratio (\textit{v}). If the critical value of \textit{v} is greater (less) than 0.26, the material is classified as ductile (brittle). From our calculated values, we report that all the pristine states of X = Cl, Br, and I, along with the pressurized states of X = I at 15 GPa and 20 GPa are ductile as their \textit{k} and \textit{v} values exceed the critical points 1.75 and 0.26. The rest are considered to be brittle. 
\par We, hereby, calculated some of the mechanical properties of these lead-free perovskites, such as anisotropic factor, Vicker hardness, melting temperature and Machinability index. These calculated figures and output are shown in Fig.S5 and from these we learn that these materials are mechanically applicable for device fabrication. 
\par To comprehend the mechanical properties of these systems, we have also calculated the sound velocities, debye's frequencies, and temperatures, which are tabulated in Tables (S6, S7 and S8).

\begin{table}[hbt!]
	\small
	\caption{\ Calculated elastic constants\textit{C$_{iv}$}, Pugh's(\textit{k}) and Poisson's\textit{(v)} ratios under different pressures in fully relaxed optimized structures of LiSnX$_3$ (X = Cl, Br, and I)}
	\label{Table 3}
	\begin{tabular*}{0.48\textwidth}{@{\extracolsep{\fill}}l|lllll}
		\hline
		& 0 GPa & 5 GPa & 10 GPa & 15 GPa & 20 GPa      \\
		\hline
		&  & & X=Cl&&\\
		\hline
		C$_{11}$ & 27.76 & 70.76  & 106.90  & 131.23 & 176.76    \\
		C$_{12}$ & 10.60  & 19.96  & 32.31  & 50.18 & 55.06   \\
		C$_{33}$ & 22.55  &  13.14 & 97.52  & 140.17 & 163.73   \\
		C$_{44}$&  2.11 & 19.22  &  28.23 & 43.18 & 62.42   \\
		B &  11.80 & 32.77  & 51.63  & 69.52 & 89.34   \\
		E & 12.29 & 55.24 & 83.69 & 107.89 & 146.96    \\
		G  & 4.63 & 22.66 & 34.03 & 43.46 & 59.94   \\
		\textit{k} & 2.55 & 1.45 & 1.52 & 1.60 & 1.50 \\
		\textit{v} & 0.33 & 0.22 & 0.23 &0.24& 0.23 \\
		\hline
		&  & & X=Br&&\\
		\hline
		C$_{11}$ & 22.32 & 65.70& 104.33 & 150.14 & 180.72 \\
		C$_{12}$ & 8.67 & 18.44 & 26.61& 38.63& 58.54 \\
		C$_{33}$ & 17.74  & 56.25 & 91.09  & 139.99 & 170.83  \\
		C$_{44}$& 1.20 & 16.30 & 28.80  & 39.41 & 44.62 \\
		B &  9.37 & 29.93  & 48.41  & 65.45 & 83.26   \\
		E & 8.69 & 50.27 & 83.32 & 110.77 & 136.37    \\
		G  & 3.23 & 20.60& 34.34 &45.47 & 55.57  \\
		\textit{k} & 2.90 & 1.45 & 1.41 & 1.44 & 1.50\\
		\textit{v} & 0.34 & 0.22 & 0.21 & 0.21 & 0.28\\
		\hline
		&  & & X=I&&\\
		\hline
		C$_{11}$ & 23.56 & 60.06  &  91.24  & 84.30 & 111.07 \\
		C$_{12}$ & 9.19  & 16.08  &  27.59 & 53.95 &31.63 \\
		C$_{33}$ & 18.89  & 52.52 &  88.77 & 147.97 & 177.10 \\
		C$_{44}$& 4.05 & 15.03& 23.64& 24.73 & 13.01\\
		B & 11.31 & 27.45  & 44.35 & 62.11 & 70.31\\
		E & 15.10 & 46.35 & 72.92 & 55.27 &56.51\\
		G  & 5.91 & 19.02 & 29.74 & 20.45 & 20.69\\
		\textit{k} & 1.91& 1.44 & 1.49 & 1.82 & 3.98 \\
		\textit{v} & 0.27 & 0.22 & 0.23 & 0.35 & 0.37 \\
		\hline
	\end{tabular*}
\end{table} 
\subsection{Piezoelectric and Electromechanical Coupling Constants}
\par Research interest in piezoelectric characteristics has increased significantly due to their green energy conversion process\cite{Adhikari2009}. Given their ability to transform mechanical stress into electrical energy, piezoelectric materials are seen as one potential use\cite{Singh2023b}. This phenomenon occurs when mechanical stress causes an atomic polarization in the material\cite{Li2022a}. Therefore, greater polarization indicates an increase in the responsiveness of piezoelectric properties\cite{King-Smith1993b}. Recent studies have shown that halide perovskites, which have broad band-gaps and non-centrosymmetric structures, exhibit high responsiveness for piezoelectricity, which depends on a material's lattice structure and interacts between a material's mechanical and electrical states\cite{Ippili2021a}. The main aim of computing the piezoelectric tensors is to generate and produce green and clean energy. Piezoelectric materials are expected to find widespread use in future and present generations for a variety of micro-electromechanical systems (MEMS)\cite{Young2016a}. Due to a lack of adequate investigations or reporting on the piezoelectric properties of the examined halide perovskites at compressive hydrostatic pressures. We subsequently calculated their piezoelectric tensors. In addition, we determined the electromechanical coupling constants to confirm the effectiveness and electrical conversion rates of these lead-free halide perovskites using the computed elastic constants, dielectric constants, permittivity of space, and the piezo-tensors.

\par Herein, it is well-known that in the absence of external field interference, the sum of the spontaneous polarization, \textit{P}$_eq$ of the equilibrium structure and the strain-dependent piezoelectric polarization generated by strain, \textit{P}$_p$ is the total macroscopic polarization (\textit{P}) of a bulk system whose expression is given as\cite{Bernardini1997b} 
\begin{equation}
P = P_p + P_{eq}
\end{equation}

\par Piezoelectric tensors can be obtained using the following expression:

\begin{equation}
\gamma_{\delta\alpha} = \dfrac{\triangle P_\delta}{\triangle \in_\alpha}
\label{Eq 15}
\end{equation}

\par where $\gamma_{\delta\alpha}$ can be computed using the QuantumATK algorithm, which uses the finite-difference method and the polarization (\textit{P}) can be acquired using the Berry-phase approximation technique\cite{Rohrlich2009b}. There are alternate ways to obtain the piezoelectric tensors, which include:
\par i) the clamped-ion e$_{ij}$ which indicates the electronic response to strain.
\par ii) a term that defines the internal strain that affects the piezoelectric polarization.
\par Therefore, the expression to obtain the piezo-tensors (e$_{ij}$) is given as 

\begin{equation}
e_{ij} = e_{ij}(0) + \dfrac{4eZ^*}{\sqrt{3}a^2} \dfrac{du}{d{\in_\alpha}}
\label{Eq 16}
\end{equation}

\par where \textit{i} and \textit{j} denoted the applied current and strain directions. Z$^*$ represents the Born effective charge, which depends solely on the polarization when the ions are displaced. The letter \textit{e} refers to the electronic charge and then $\in_\alpha$ is the macroscopic applied strain.

Our obtained piezoelectric tensors are tabulated in Tables S10, S11, and S12 respectively. Here, the underinvestigated halide perovskites tensors are generated along three directions, i.e., \textit{x}, \textit{y}, and \textit{z} axes in six different strains: \textit{ xx, yy, zz, yz, xz, and xy}. Multiple deformation orientations ($\eta$) cause the structure to be disrupted, causing the negative cloud of electrons to orient towards the positive nucleus, generating visible deformation. The electric field produced by this small distance between them causes polarization, which can be altered to produce unique piezoelectric characteristics. As can be seen from Tables S10-S12, the piezo-tensors changes with increasing pressures. The highest tensor response obtained for each compound is tabulated in Table \ref{Table 4} LiSnCl$_3$ (\textit{e}$_{31}$ = 0.69 C/m$^2$) has been observed at 5 GPa along the \textit{z}-axis under a positive \textit{xx}-strain. Furthermore, we observed the maximum tensor for LiSnBr$_3$ (\textit{e}$_{33}$ = 20.7 C/m$^2$) at 15 GPa along \textit{z}-axis under the positive \textit{zz}-strain. Furthermore, for LiSnI$_3$, the highest piezo-response has been observed to be \textit{e}$_{33}$ = -0.64 C/m$^2$ at 20 GPa along \textit{z}-axis whose tensor arises under the \textit{zz}-strain. Each of these piezoelectric response values under investigation is substantially greater than the usual $\alpha$-quartz piezo-response coefficients (0.17 C/m$^2$) obtained by Bechmann \textit{et al.}\cite{Bechmann1958c} and temperature-dependent work (0.01 C/m$^2$) performed by Tarumi \textit{et al.}\cite{Tarumi2007b}. This high value of piezoelectric constant in LiSnX$_3$ surpassed our previous theoretical results of 12.48  C/m$^2$ in the same perovskite family\cite{Celestine2024d}. As can be seen from the electron distribution in Fig.\ref{Fig.4 EDD} and S13, we observed that the atoms are located more in the x-, and y- axes which means cancellation of charges are higher in this region or directions. At the z-axis, there are fewer numbers of atoms and when strain is given, the responsiveness of charge polarization is more without canceling each other, which yielded high piezoelectric responses in the X = Br series. The high piezo-response obtained at this particular 15 GPa may be due to the orientation of atoms under this specific applied pressure, where the atoms are arranged to have an excellent range for polarization.  We can therefore conclude that the halide perovskites under investigation using this pressure-induced technique yield a remarkable piezoelectric response, confirming the potential for use of the materials in ferroelectric and piezoelectric applications.

\par The efficiency rate of a piezoelectric material that transforms mechanical stress into electrical energy is determined by an electromechanical coupling constant (\textit{k$_{ij}$}). In this paper, the subscript "\textit{i}" denotes the applied electrode direction, while the "\textit{j}" subscript denotes the path along the applied mechanical tension. According to the supplier's piezoelectric performance, the \textit{k} values represent the theoretical system's maximum values\cite{Hassani2021}. Consequently, the equation provided can be used to calculate the efficiency values. \cite{Roy2012c}

\begin{equation}
\textit{k}_{ij} = \dfrac{|\textit{e}_{ij}|}{\sqrt{C_{ij}\epsilon_{fs}\epsilon_0}}
\label{Eq 17}
\end{equation}

\par From the above equation, the term \textit{k$_ij$} represents the electromechanical coupling coefficients, \textit{e$_{ij}$} denotes the calculated piezoelectric coefficient (Cm$^{-2}$), C$_{ij}$ being the elastic constant (10 $^9$ Pa), $\epsilon_{0}$ represents the static dielectric constant at the zero stress level and $\epsilon_{fs}$ is the permittivity of space. We employed perturbed density functional theory (DFT), as implemented in QuantumATK, to perform a linear response computation and determine the electronic dielectric constants.

\begin{table}[hbt!]
	\small
	\caption{\ Calculated values of the piezo-electromechanical coupling constants for LiSnX$_3$ (X=Cl, Br and I) where e$_{ij}$ represents the piezoelectric coefficients and k$_{ij}$ being the electromechanical coupling coefficient.}
	\label{Table 4}
	\begin{tabular*}{0.48\textwidth}{@{\extracolsep{\fill}}l|l|ll}
		\hline
		\textbf{X} &Pressure& \ \textit{e$_{ij}$} & \textit{k$_{ij}$} \\
		\hline
		\textbf{Cl}     &5 GPa& 0.69   &1.29 \\
		\textbf{Br}     &15 GPa&  20.7   &30.06\\
		\textbf{I}     &20 GPa&0.64   & 1.36 \\
		\hline
	\end{tabular*}
\end{table}

\par  In Table \ref{Table 4}, we present the electromechanical coupling constants by calculating the highest piezoelectric constants from each lead-free halide perovskite under investigation. Using these calculated data, we can conclude that among the compounds studied, pressure has such a great influence on the properties within the materials, and thus confirmed that the Br-atom-containing compound with 15 GPa-induced pressure has produced the highest electromechanical coupling efficiency. Regardless of whether a larger \textit{k$_{ij}$} value indicates greater energy conversion efficiency, the material may make a great electromechanical transducer between the defined electric and elastic channels when the coupling factor values are close to unity. \cite{Wu2005a}

\section*{Conclusions}
\par In this work, the eco-friendly halide-based perovskites LiSnX$_3$ (X = Cl, Br, and I) are investigated and examined under compressive isotropic pressures (0 to 20 GPa). In this paper, we report that the inducing pressures influence the properties of the compounds. The stability of the compounds was validated using formation energy, MD simulation, and elastic constant calculations. We also observed the lattice parameters and \textit{E$_g$} received decreases under applied pressures, and all systems considered are semiconductors. Under the influence of pressure, the halide perovskites exhibited considerable dielectric polarization, yielding piezoelectric constants. From X = Br at 15 GPa, we obtained the highest piezo-response with 20.7 C/m$^2$ which has the electromechanical coupling constant of 30.06. Therefore, we conclude that the examined materials are promising and capable of green energy harvesting via piezoelectric applications.

\section{Acknowledgments}
The research is partially funded by the Ministry of Science and Higher Education of the Russian Federation as part of the World-Class Research Center program: Advanced Digital Technologies (contract No. 075-15-2022-312 dated 20.04.2022).\\	
A. Laref acknowledges support from the "Research Center of the Female Scientific and Medical Colleges",  Deanship of Scientific Research, King Saud University.

\section*{Author contributions}
\textbf{Celestine Lalengmawia:} Formal analysis, Visualization, Validation, Literature review, Performed Calculation, Writing-original draft, writing-review \& editing.\\
\textbf{R. Zosiamliana:} Formal analysis, Visualization, Validation, writing-review \& editing. \\
\textbf{Bernard Lalroliana:} Formal analysis, Visualization, Validation, writing-review \& editing. \\
\textbf{Lalhum Hima:} Formal analysis, Visualization, Validation, writing-review \& editing. \\
\textbf{Shivraj Gurung:} Formal analysis, Visualization, Validation, writing-review \& editing. \\ 
\textbf{Lalhriat Zuala:} Formal analysis, Visualization, Validation, writing-review \& editing. \\
\textbf{Lalmuanpuia Vanchhawng:} Formal analysis, Visualization, Validation, writing-review \& editing. \\ 
\textbf{Amel Laref}:Formal analysis, Visualization, Validation, writing-review \& editing. \\	
\textbf{A. Yvaz:}Formal analysis, Visualization, Validation, writing-review \& editing. \\ 
\textbf{D. P. Rai:} Project management, Supervision, Resources, software, Formal analysis, Visualization, Validation, writing-review \& editing. 

\section*{Conflicts of interest}
There are no conflicts to declare.

\section*{Data availability}
The data that support the findings of this study are available from the corresponding author upon reasonable request.

\nocite{*}
\bibliography{apssamp.bib}

\end{document}